\documentclass[a4paper,twocolumn]{scrartcl}

\usepackage{artsci}
\usepackage{color}
\usepackage{tabularx}

\JOURNALtrue

\JAPANESEfalse

\newcounter{FirstPage}
\newcommand{\asheader}{}

\newcommand{\aspagenum}{\thepage}

\usepackage{amsmath, amssymb}	
\usepackage{graphicx}	
\usepackage{float}		
\usepackage{url}		

\graphicspath{{fig/}} 

\etitle{
Lottery and Sprint Arcade: Enabling Player-Driven Game Editing with Generative AI
}

\eauthor{
Maya Grace Torii\(^{1)}\)
Takahito Murakami\(^{1)}\)
Yoichi Ochiai\(^{2)}\)
}

\eaffiliation{
1) Graduate School of Comprehensive Human Sciences, University of Tsukuba\\
2) R \& D Centre for Digital Nature, University of Tsukuba
}

\email{\{toriparu, takahito\} (at) digitalnature.slis.tsukuba.ac.jp \\}

\eabstract{
Large language models (LLMs) are shifting game generation from offline automation toward play-driven modification through natural-language interaction. In this work, we present a play-driven game editing system that enables players to modify a retro Space Invaders -- style arcade game through voice-based natural-language commands during play. Spoken instructions are interpreted by an LLM and translated into structured updates of internal configuration parameters, allowing iterative play $-$ edit -- feedback cycles in an invader-style game environment without exposing underlying system details. The game includes approximately 100 editable configuration fields controlling mechanics, visuals, interaction patterns, and audio behavior, enabling gameplay transformation through incremental parameter changes.
To investigate how users experience play-driven AI-mediated editing (RQ1) and how emergent editing patterns relate to variations in player experience (RQ2), we conducted a user study combining subjective evaluations, workload measures, and log-based analysis of editing behavior. Participants were able to modify gameplay with generally positive experiences and moderate workload, and interaction outcomes did not strongly depend on prior programming experience. Editing-log analysis revealed distinct experiential tendencies: adjustments to immediately perceptible parameters were associated with higher usability, whereas edits affecting core gameplay structures were more closely associated with enjoyment. Post-session reflections further identified diverse editing strategies, including exploratory experimentation, goal-driven structural modification, and iterative parameter tuning.
These findings demonstrate that voice-driven editing can support accessible, play-driven human -- AI co-creation within a structured invader-style arcade game environment.
}

\begin{document}
\maketitle
\thispagestyle{aspagestyle}
\noindent
\makebox[\columnwidth][l]{%
  \makebox[\textwidth][l]{%
    \raisebox{-10mm}[0pt][0pt]{%
      \parbox{0.9\textwidth}{%
        \raggedright
        \scriptsize
        \textit{Author-prepared manuscript. The version of record was published in}\\
        \textit{The Journal of the Society for Art and Science},
        Vol.~25, No.~2, pp.~12:1--12:14, 2026.
      }%
    }%
  }%
}

\clearpage

\section{Introduction}
Digital game generation has been explored for many years through procedural content generation (PCG), algorithmic content generation (ACG), and machine learning -- based approaches. These studies have demonstrated that games, levels, and rules can be produced by computational systems~\cite{shaker2016procedural, piette2019ludii}.
More recently, large language models (LLMs) have introduced a new paradigm. Instead of generating complete games offline, LLM-based systems allow users to interact through natural language and iteratively modify content during use. LLMs shift game generation from offline automation to play-driven modification and co-creation through natural-language interaction~\cite{maleki2024procedural, zand2025democratizing, sudhakaran2023mariogpt}. In practice, AI-mediated editing may not simply reshape an existing game but lead to new forms of play or entirely different game experiences~\cite{charity2020baba}. 

In this work, we present a play-driven game editing system that enables players to modify a retro Space Invaders -- style arcade game through voice-based natural-language interaction during play. The system builds on our previous “Lottery and Sprint” method~\cite{torii2023lottery}, a board game creation methodology that combines an LLM-based agent with the structured Design Sprint framework to support human -- AI co-creation. The method was designed to make game design accessible to non-expert users by guiding iterative idea generation and refinement through AI-assisted workflows. Although Lottery and Sprint demonstrated that beginners could collaboratively design games with AI support, the process remained slow because each iteration required reading rules and completing full play sessions. To enable faster experimentation, we extend this approach to a video game context, where changes can be experienced immediately during play. Situated at the intersection of human--computer interaction and interactive game experience design, this work investigates how voice-based natural language interfaces can support play-driven modification accessible to non-expert users.

In this system, players issue voice commands to modify a retro arcade -- style game during play, and the LLM updates internal configuration parameters in response to each play cycle. The game contains approximately 100 editable configuration fields controlling mechanics, visuals, interaction patterns, and audio behavior, enabling gameplay to be transformed through incremental parameter changes rather than external modifications.
To investigate the usability and creative potential of AI-mediated game editing, we conducted a user study combining subjective evaluations of user experience and workload with analysis of editing logs and the diversity of generated gameplay outcomes. RQ1: How do users interact with and experience play-driven AI-mediated game editing, and does prior programming experience influence interaction patterns or perceived experience? RQ2: How do different editing patterns emerging through iterative natural-language editing relate to variations in player experience across trials?

The contributions of this work are threefold. First, we present a play-driven AI-mediated game editing system that enables natural-language modification of a retro arcade game. Second, provide empirical findings suggesting that such editing can be accessible without strong dependence on prior programming experience. Third, we demonstrate the diversity of gameplay outcomes enabled by iterative natural-language editing.

\section{Related Work}

\subsection{LLMs for Game Content Generation}
Traditional procedural content generation (PCG) approaches surveyed by Shaker et al.~\cite{shaker2016procedural} relied on algorithmic methods such as rule-based systems, generative grammars, and machine learning to produce game levels, rules, and assets without direct player input. More recently, LLMs have introduced natural language as a medium for specifying generated content. MarioGPT~\cite{sudhakaran2023mariogpt} demonstrated that a fine-tuned language model could generate playable Super Mario Bros.\ levels from short text prompts, establishing a direct mapping between natural language descriptions and game structure. GAVEL~\cite{todd2024gavel} extended this further, combining evolutionary algorithms with LLMs to automatically generate complete board games with novel rules and mechanics.

Alongside generation quality, the accessibility of authoring interfaces has received increasing attention. LLMaker~\cite{gallotta2024llmaker} proposed a chat-based level design interface that uses the function-calling capabilities of LLMs to produce consistent, rule-compliant game content from natural language input. Whitehead et al.~\cite{whitehead2025conversational} explored conversational interaction with procedural generators, demonstrating that LLMs can support mixed-initiative game world generation through dialogue. Miralvand et al.~\cite{zand2025democratizing} showed that generative AI tools can lower the barrier to game modification, enabling players without technical backgrounds to iteratively create content through natural language interaction.

A common thread across these works is that content generation and editing occur outside of active play: content is created or modified before or between sessions, and the player experiences the results in a separate play phase. The present work builds on this iterative authoring paradigm, extending it to a play-driven context in which users edit in response to what they experience during active play, rather than refining a design artefact outside of play.

\subsection{In-Play and Interactive Game Editing Tools}
Several systems have explored integrating editing capabilities into active gameplay. Tanagra~\cite{smith2010tanagra}, a foundational mixed-initiative level design tool, enabled real-time collaboration between a designer and an AI that automatically generated level geometry in response to designer actions. Jennings et al.~\cite{jennings2024gromit} introduced GROMIT, a runtime behavior generation system for Unity that compiles LLM-generated game behaviors during play without developer intervention, achieving an 85\% success rate in a controlled evaluation. Both systems demonstrate the technical feasibility of editing during or alongside active gameplay. However, both were designed for and evaluated with professional users: Tanagra targeted level designers familiar with spatial design concepts, while GROMIT targeted game developers. The unintended downstream gameplay changes reported in 4--5\% of GROMIT generation events, which the authors note require additional guardrail systems, further motivate a more conservative approach for non-expert users.

van Rozen~\cite{vanrozen2025livegame} proposed Live Game Design, a rapid prototyping approach that enables simultaneous editing and playtesting through mini-cycles, targeting designers across skill levels. Oswald et al.~\cite{oswald2014realAR} demonstrated a real-time video game level editor in a spatial augmented reality environment, allowing users to incorporate everyday physical objects into game content during play. Charity et al.~\cite{charity2020baba} explored collaborative mixed-initiative level design in Baba Is Y'all, where game rules themselves are manipulable gameplay objects. These works collectively reflect a growing interest in collapsing the boundary between playing and authoring, which the present work extends to non-expert players through voice-based natural language interaction.

\subsection{Natural Language Interfaces for Non-Expert Users}
A central design goal of the present system is accessibility for users without programming or game development experience. Research on end-user programming with LLMs has identified a persistent \textit{abstraction gap} between users' high-level intent and the formal representations a system can act upon~\cite{jiang2022whatwants}. Users tend to describe desired outcomes rather than procedural steps, preferring goal statements such as ``make it more exciting'' over technical specifications~\cite{how2024humans}, and the LLM must bridge this gap by mapping natural language to concrete operations. In the present system, this mapping is from spoken commands to structured JSON configuration updates, and the voice-based modality is chosen precisely because it requires no knowledge of the underlying parameter structure.

Miralvand et al.~\cite{zand2025democratizing} demonstrated that generative AI tools can democratize game modification, enabling players to iteratively create and refine game content with minimal input. Panchanadikar and Freeman~\cite{panchanadikar2024solo} examined how indie game developers envision the role of generative AI in creative workflows, finding both enthusiasm for accessible AI assistance and concern about the displacement of creative agency, a tension also present in player-facing tools. Liebers et al.~\cite{liebers2024compgen} evaluated control strategies and input modalities for AI-generated virtual environments, finding that voice input combined with step-by-step generation supported a sense of user control and manageable cognitive load relative to more direct manipulation approaches.

\subsection{Human-AI Co-Creation in Games and Creative Practice}
The present work engages with a growing body of research on human-AI co-creation in game and creative contexts. Guzdial et al.~\cite{guzdial2019moraimaker} developed Morai Maker, an AI-driven game level editor for Super Mario Bros.-style games, and found through studies with over one hundred participants that designers adopt diverse orientations toward an AI collaborator --- framing the AI variously as a Friend, Collaborator, Student, or Manager depending on their creative goals and degree of desired control. These roles reflect different balances of agency between human and AI, a dimension directly relevant to the present work: participants in our study similarly exhibited distinct orientations, ranging from exploratory engagement --- treating the AI's unpredictability as a creative resource --- to structural direction, in which the AI served as an instrument for realizing specific aesthetic or gameplay visions.

Moruzzi and Margarido~\cite{moruzzi2024usercentered} proposed a user-centered framework for human-AI co-creativity that identifies key dimensions of agency and control modulation in co-creative processes, underscoring the importance of allowing users to adjust the balance between AI autonomy and human direction. Charity et al.~\cite{charity2023inksplotch} demonstrated, through the ``ink splotch effect,'' how LLM-generated content can function as a creative catalyst that stimulates unexpected design directions rather than simply fulfilling explicit requests. In a related vein, Charity et al.~\cite{charity2023langreality} developed a co-creative storytelling game in which natural language is the medium through which players and AI jointly construct game reality, emphasizing the generative potential of language as a creative interface.

The present system is a direct extension of our prior work, Lottery and Sprint~\cite{torii2023lottery}, bringing the human--AI collaborative game design approach demonstrated there into a video game context to enable faster iterative cycles through play-driven editing.

\section{Implementation}
\begin{figure*}[thbp]
    \centering
    \includegraphics[width=1\linewidth]{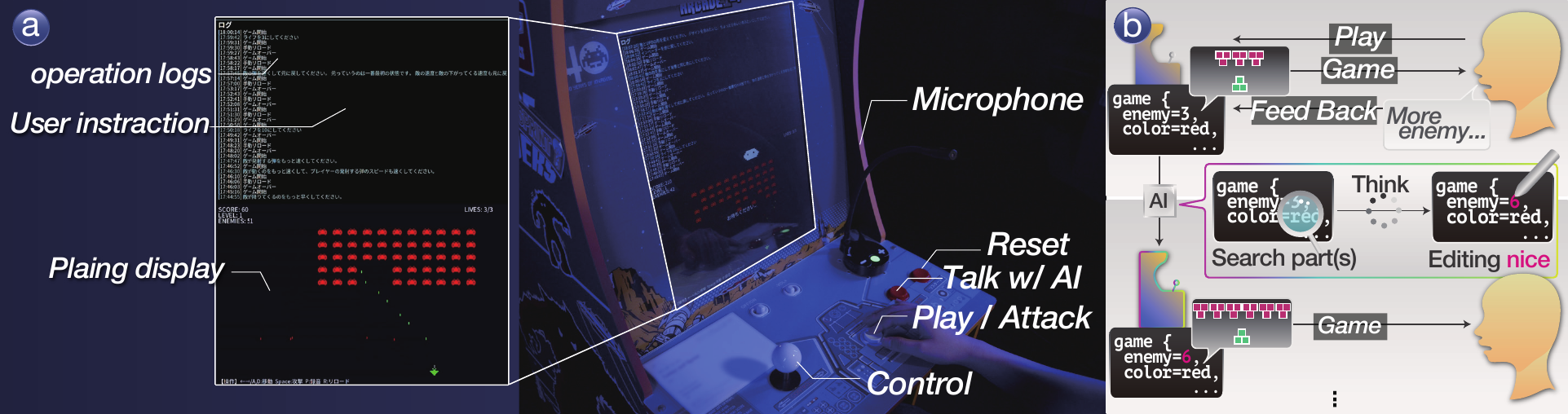}
    \caption{System overvie: (a) gameplay interface and (b) retro arcade-style voice-controlled setup, and (c) LLM-guided configuration update workflow. The modified configuration is immediately reflected in the game, enabling iterative play -- edit -- feedback cycles.}
    \label{fig:hardware}
\end{figure*}

\subsection{System Overview and Gameplay Interaction}
The system runs on a modified Arcade1Up cabinet retrofitted with a Raspberry Pi 5, and using OpenAI Whisper and GPT-4o for LLM functions. The cabinet retains its original gameplay functions and physical interface design (Figure~\ref{fig:hardware}(a),(b)). A retro arcade format was selected because its widely recognized gameplay reduces the learning curve for participants regardless of gaming background, and its simple rule structure makes the configuration space tractable for LLM-based editing. A microphone was added as an additional input device, and the existing “Player Select” button was repurposed as a dedicated “Voice Input” button to toggle voice command mode. Voice input was selected as the primary editing modality because speech allows users to express editing intent in natural language without learning specialized syntax or interface conventions.

Gameplay follows the standard Space Invaders format. During play, the user presses the voice input button to issue spoken commands (e.g., “make it faster” or “change bullet color”), which are processed via the Whisper API for speech recognition and interpreted by GPT-4o to generate updates to the game’s configuration file. When an update succeeds, the game resets and restarts with the modified parameters. This reset-based design was chosen to preserve a clear before-and-after comparison for non-expert users: applying edits mid-session would make it difficult to attribute gameplay changes to specific commands, particularly for participants without prior programming experience. Configuration values and available options are not exposed to the player; changes are instead perceived indirectly through subsequent gameplay behavior, visuals, or sound.

The configuration update process is implemented as a language model -- guided pipeline that converts voice commands into structured edits of a JSON-based configuration file (Fig.~\ref{fig:hardware}(c)).
After transcription by Whisper, the text command is sent to gpt-4o-2024-08-06 together with the current game configuration and a predefined JSON schema defining editable parameters (Table~\ref{tab:core_configuration_fields}).

Configuration updates are applied in two stages: \emph{plan} and \emph{action}. In the plan stage, the model maps the user instruction to a structured list of proposed parameter changes. Each entry includes the target field, its current value, the proposed new value, the rationale for the change, and its expected effect on gameplay. This stage is constrained by a prompt that enforces a fixed output structure (see Supplementary Materials, Listing~1), enabling reliable parsing and inspection. In the action stage, planned changes are converted into atomic JSON patch operations that modify the game configuration (e.g., \texttt{game.enemies.speed: 1.2 → 1.6}) (Fig.~\ref{fig:hardware}(c)).

\begin{table}[t]
\footnotesize
\setlength{\tabcolsep}{3pt}
\centering
\sloppy
\caption{Core configuration fields targeted by plan/action updates.}
\label{tab:core_configuration_fields}
\begin{tabular}{@{}l >{\raggedright\arraybackslash}p{2.8cm} p{3.4cm}@{}}
\hline
\textbf{Category} & \textbf{Field(s)} & \textbf{Role} \\
\hline
Appearance   & color, pixel\_size, pattern                        & Pixel-art sprite definitions. \\
Player       & speed, fire\_rate, initial\_lives, max\_lives      & Player movement, firing, and life limits. \\
Enemies      & rows, cols, speed, shoot\_chance, health\_by\_row  & Enemy layout, movement, attack behavior, and durability. \\
Bullets      & bullets                                            & Bullet speed (player / enemy). \\
UFO          & spawn\_probability, min\_spawn\_interval, speed, points & Controls rare enemy timing and reward. \\
Barriers     & barriers                                           & Count, size, and durability. \\
Game limits  & game\_settings                                     & Max simultaneous enemy bullets. \\
Level        & enemy\_speed, shoot\_chance                        & Difficulty scaling across levels. \\
Power-ups    & powerups                                           & Drop rate, effects, and appearance. \\
Sounds       & sounds                                             & Sound file paths and volume. \\
General      & title, background\_color                           & Title and background color. \\
\hline
\end{tabular}
\end{table}

\subsection{Logging and Instrumentation}
All voice-driven interactions are recorded in structured logs for analysis and evaluation, including user instructions, configuration updates, and associated system outputs (Section~4.2).
\section{User Study}
We examined how players experience play-driven editing via voice commands of gameplay parameters via LLM. We focused on subjective usability impressions during repeated play--edit cycles and tested whether these impressions relate to (i) editing amount, (ii) edit \emph{type} observed in logs, and (iii) prior programming experience.

\subsection{Participants and Ethics}
We recruited 21 adults from a university campus via posters and online boards (12 men, 9 women; age: $M{=}26.71$, $SD{=}7.98$, range{=}21--52). To avoid bias toward experts, we targeted a wide range of technical backgrounds. Participants provided written informed consent. The study was approved by the University of Tsukuba Ethics Committee (ID: 25-47). No personally identifiable information was recorded.

\subsection{Procedure}
Prior to each session, participants received an explanation of the study procedure, followed by a tutorial covering standard Space Invaders gameplay rules, physical button controls, and voice input operation. Participants were provided with two reference documents available for consultation throughout the session. One was a one-page overview listing the core configuration categories (corresponding to Table~\ref{tab:core_configuration_fields}), and the other was a full configuration reference describing all editable fields in detail (Supplemental Material). The transcription of each spoken command was displayed on screen, allowing participants to confirm what had been recognized; however, the specific configuration fields modified and their updated values were not shown, so participants perceived the effects of edits solely through subsequent gameplay behavior.
Each participant completed one session with five gameplay--editing trials. After a brief tutorial and background questionnaire, each trial began from a reset state. During play, participants could issue voice commands at any time; each trial required at least two successfully applied edits and allowed iterative edit--play cycles until voluntary termination. After each trial, participants completed a five-item (7-point Likert) trial questionnaire. After the fifth trial, they completed post-session questionnaires (UEQ, NASA-TLX, and study-specific items).

\subsection{Measures}
We collected (1) demographics and background, (2) a repeated five-item trial questionnaire (editing ease, controllability, enjoyment, confusion [reverse], intention--behavior alignment), and (3) post-session questionnaires: UEQ~\cite{laugwitz2008ueq} (Japanese version; six subscales), NASA-TLX~\cite{hart1988nasatlx,haga1996nasatlx} (Japanese adaptation; six dimensions), and eight study-specific Likert items about the interface. Full wording of the additional post-session items is provided in S.3 Post-session questionnaire items (Supplementary Material).

\subsection{Analysis}
Analyses were conducted in Python and R\footnote{Python: \texttt{scikit-learn} for preprocessing and PCA; R: \texttt{lme4} for linear mixed-effects models, \texttt{lmerTest} for fixed-effect testing, and \texttt{rstatix} for Wilcoxon tests with multiple-comparison correction.}. Trial-level ratings ($21 \times 5 = 105$) were analyzed using linear mixed-effects models with participant random intercepts. Editing amount (number of successfully validated configuration-field updates) and trial order were treated as fixed effects to examine interaction intensity and learning effects.

To capture edit type, logs were encoded as TF-IDF features based on second-level configuration paths and reduced via PCA. Raw frequency counts were not used because the number of log entries per trial scales with editing volume (M = 6.81, SD = 4.58 operations per trial); TF-IDF normalizes within each trial, allowing the analysis to capture editing focus independently of editing quantity. The resulting components (PC1, PC2) were entered as fixed effects in mixed-effects models to test associations with subjective ratings.

For post-session outcomes (NASA-TLX, UEQ, study-specific items), participants were grouped by programming experience (Pro $n{=}6$, Amateur $n{=}8$, None-EX $n{=}7$) and compared using pairwise Wilcoxon tests with multiple-comparison correction.
\section{Result}
\begin{figure*}
    \centering
    \includegraphics[width=1\linewidth]{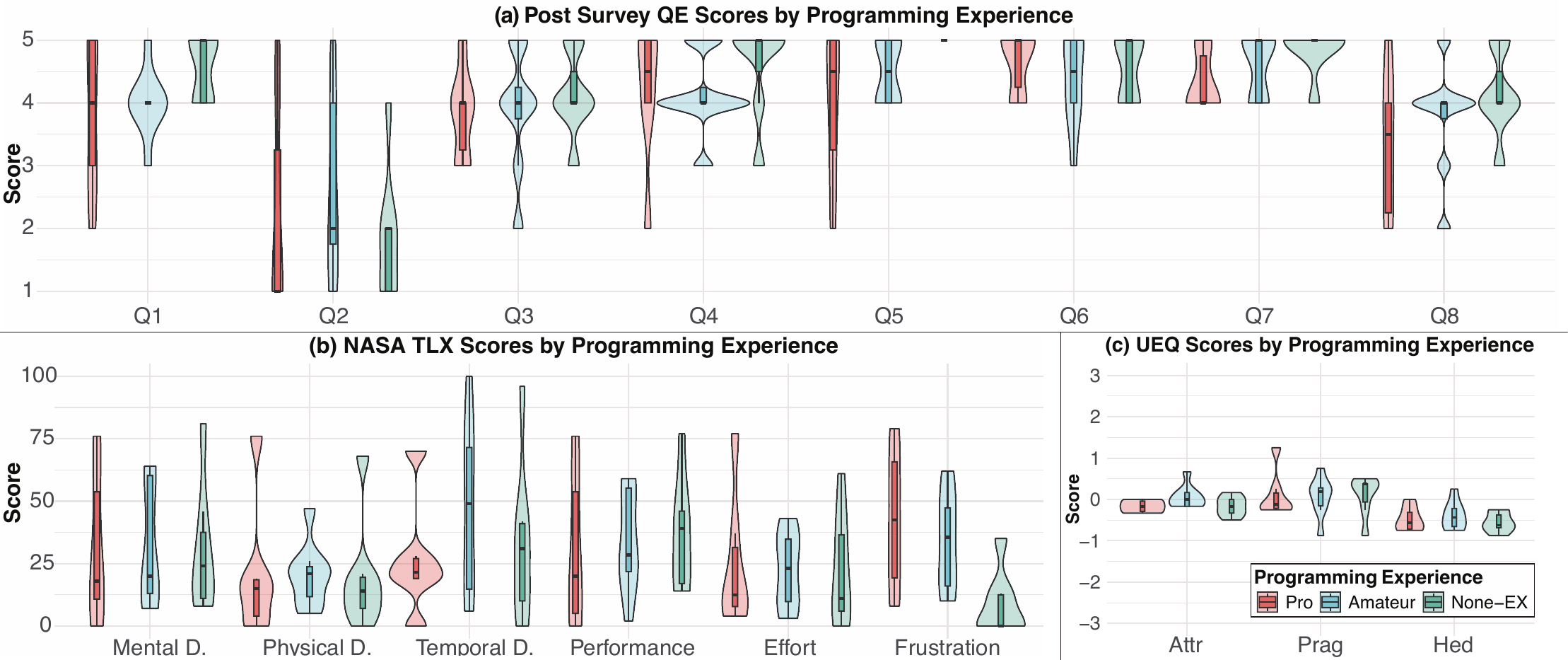}
    \caption{
    Subjective evaluation results.
    (a) NASA-TLX subscales and Overall Workload by programming expertise.
    (b) Post Survey item means (Q1--Q8).
    (c) UEQ scales (Attractiveness, Pragmatic, Hedonic).
    Error bars indicate standard deviation.}
    \label{fig:subjective}
\end{figure*}

\paragraph{System Reliability.}
Across all 105 trials, participants issued a total of 715 voice commands. Of these, 694 (97.1\%) resulted in successfully validated configuration updates. The remaining 21 (2.9\%) were rejected without modifying the game state, ensuring no unintended side effects. These figures indicate that the LLM-based editing pipeline operated reliably throughout the study.

\subsection{RQ1: User Interaction and Experience During AI-Mediated In-Play Editing}

\paragraph{NASA-TLX (Workload).}
As shown in Fig.~\ref{fig:subjective}(a), NASA-TLX scores remained within a moderate range across all participants. Mean values for all six subscales (Mental, Physical, Temporal, Performance, Effort, and Frustration) were generally between 20 and 40 points. For example, Mental Demand was similar across groups (Pro: M=31.00, SD=31.53; Amateur: M=32.25, SD=24.88; None-EX: M=30.00, SD=26.17), and Physical Demand remained low (approximately 19--21 points). Overall Workload scores were also moderate (Pro: M=27.17, SD=27.30; Amateur: M=25.62, SD=11.29; None-EX: M=10.71, SD=11.34). These results indicate that participants did not experience in-play editing as excessively demanding.

\paragraph{Post Survey and UEQ.}
As shown in Fig.~\ref{fig:subjective}(b,c), Post Survey ratings were generally positive, with most item means near 4 on a 5-point scale. Participants reported that editing was easy to perform, interaction felt natural, and editing improved the gameplay experience. UEQ scores showed mid-to-high ratings for Attractiveness, Pragmatic Quality, and Hedonic Quality. Together, these findings indicate that AI-mediated editing during play was broadly well received.

\paragraph{Exploratory comparison across programming experience.}
Participants were grouped by programming experience into Pro (n=6), Amateur (n=8), and None-EX (n=7). As shown in Fig.~\ref{fig:subjective}, pairwise Wilcoxon tests with multiple-comparison correction revealed no statistically reliable differences between groups for NASA-TLX, Post Survey items, or UEQ scales (all corrected $p$ values non-significant; $|r|$ ranged from 0.02 to 0.17, 95\% CIs spanning zero).

\subsection{RQ2: Editing Patterns and Their Relationship to Player Experience}

\paragraph{Trial-level relations between editing patterns and player experience.}
\begin{figure*}
    \centering
    \includegraphics[width=1\linewidth]{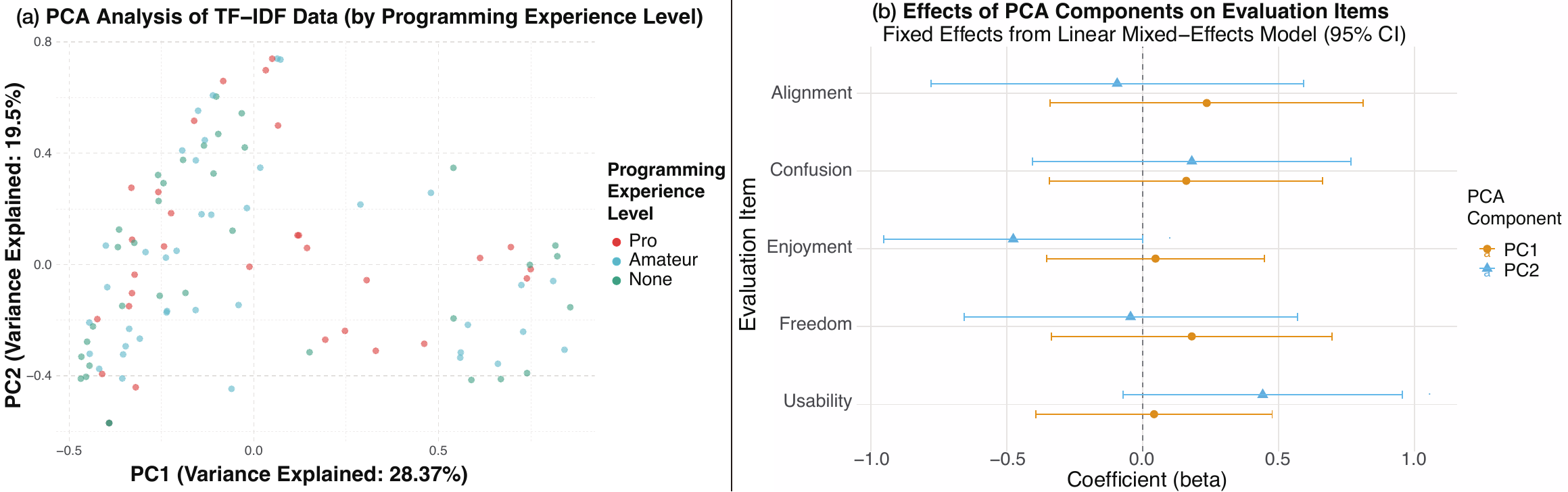}
    \caption{
   Editing-log analysis.(a) PCA scatter plot of trial-level editing logs (PC1 vs. PC2), colored by programming experience.(b) Fixed-effect estimates from linear mixed-effects models relating PC1 and PC2 to trial-level subjective ratings (95\% CI).}
    \label{fig:pca_lmm}
\end{figure*}
To examine how different editing patterns relate to experiential variation, we analyzed editing logs using TF--IDF features followed by principal component analysis (PCA) and linear mixed-effects models (LMMs). PC1 and PC2 explained 28.37\% and 19.50\% of the variance, respectively (47.86\% cumulative), providing a low-dimensional representation of editing tendencies at the trial level rather than participant attributes.

\begin{figure}
    \centering
    \includegraphics[width=\linewidth]{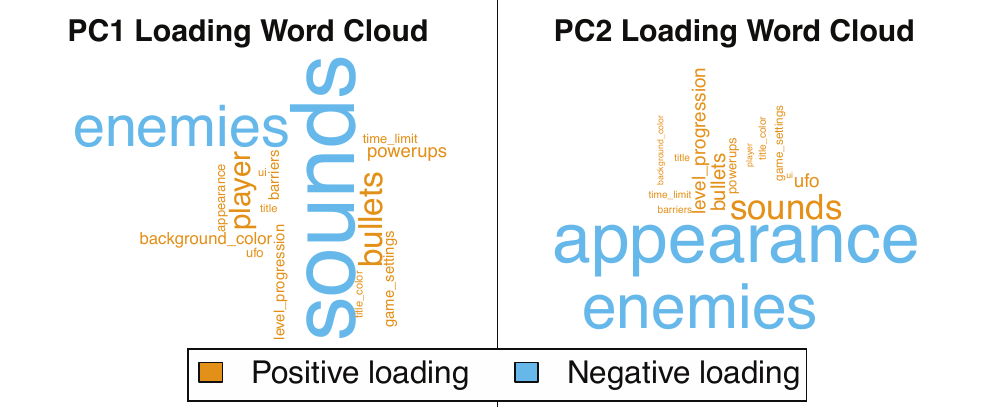}
    \caption{Word clouds of the highest-loading configuration terms.}
    \label{fig:wordcloud}
\end{figure}

Based on loading patterns, PC2 contrasted edits focused on immediately perceptible adjustments (e.g., appearance-related parameters, bullets, or UFO elements) with edits affecting core gameplay behaviour (e.g., enemies, sounds, and progression systems). PC1 reflected a distinction between experiential or presentational adjustments (e.g., sounds and powerups) and modifications targeting core interaction elements such as enemies, bullets, and player behaviour (Fig.~\ref{fig:wordcloud}). The semantic labels assigned to PC1 and PC2 are interpretive descriptions derived from inspection of the highest-loading configuration terms (Figure~4) and are not intended as confirmed categorical distinctions; given the limited sample size and potential dependence across trials, these labels serve as heuristic descriptors to facilitate discussion rather than theoretically derived or stable constructs.

Linear mixed-effects analyses examined relationships between these editing dimensions and trial-level subjective ratings collected after each iteration. PC2 showed trend-level associations with usability ($\beta=0.442$, 95\% CI $[-0.078,\ 0.961]$, $p=0.095$) and enjoyment ($\beta=-0.475$, 95\% CI $[-0.957,\ 0.006]$, $p=0.053$), corresponding to small-to-medium magnitude effects per standard deviation of PC2. These suggest that trials characterized by relatively more immediately perceptible adjustments tended to be experienced as easier to use but not necessarily more enjoyable, whereas edits targeting deeper gameplay structures showed the opposite tendency. The confidence intervals span zero, indicating substantial uncertainty under the current sample size. No other reliable associations were observed for the remaining components or rating scales.

The PCA scatter plot in Fig.~\ref{fig:pca_lmm}(a) shows substantial overlap across programming-experience groups, indicating that editing patterns were not strongly structured by prior programming expertise. Accordingly, PC1 and PC2 are interpreted as dimensions of editing behaviour emerging from interaction dynamics during iterative editing rather than differences between participant groups.

\paragraph{Qualitative interpretation of editing strategies.}

To complement the log-based quantitative analysis, we qualitatively analyzed post-session written reflections collected from all participants. Coding was performed in two stages: initial descriptive coding by two researchers followed by iterative grouping into higher-level themes through discussion and consensus. Three themes emerged: \textit{Exploratory Editing}, \textit{Structural Editing}, and \textit{Iterative Tuning}.

\textit{Exploratory Editing} described participants issuing commands without fixed expectations and engaging with unpredictable outcomes. Participants reported exploring system affordances and testing the limits of what could be modified. One participant noted that ``it is difficult to understand how the play changes depending on the result of voice input, but I was able to find fun in the unexpectedness'' (P16, None-EX), while another described finding interest in ``seeing the unpredictable changes'' (P5, Amateur). Participants frequently framed editing as an exploratory process of discovering system capabilities.

\textit{Structural Editing} involved participants pursuing specific aesthetic or gameplay goals through iterative modification. Participants described attempts to transform gameplay toward particular genres or styles, such as approximating a bullet-hell experience (P6, Pro) or creating a psychedelic audiovisual style (P7, Pro). Several participants expressed a sense of creating a new game through editing, with one stating that the process felt like ``creating my own new game'' (P16, None-EX). Comments also reflected awareness of system constraints, such as difficulty adding entirely new mechanics beyond configuration parameters (P21, Pro; P20, Pro).

\textit{Iterative Tuning} reflected a more systematic approach in which participants adjusted parameters incrementally or explored extreme values to understand system behaviour. Participants described testing boundary values to observe changes and refine gameplay characteristics, with one noting that extreme settings were used to understand robustness and system limits (P6, Pro). Some participants framed the system as useful for fine-tuning gameplay variables during development rather than primarily for experiential exploration (P20, Pro).

These qualitative themes provide contextual grounding for the editing patterns identified in the PCA analysis by describing distinct ways participants engaged with in-play editing.

\paragraph{Documentation reference patterns.}

\begin{figure}[h]
    \centering
    \includegraphics[width=\linewidth]{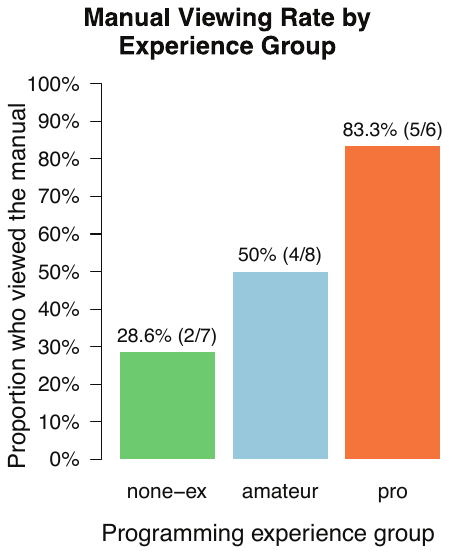}
    \caption{Proportion of participants who consulted the provided documentation, by programming experience group. Consultation rates decreased with lower experience (Pro: 83.3\%, Amateur: 50.0\%, None-EX: 28.6\%).}
    \label{fig:manual_reference}
\end{figure}

We additionally examined whether participants referred to the provided documentation based on keyword-based coding of post-session comments (e.g., ``manual'', ``reference'', ``documentation''). As shown in Fig.~\ref{fig:manual_reference}, references to the documentation were more frequently observed among participants with prior programming experience (83.3\%), while fewer mentions were observed in the Amateur (50.0\%) and None-EX (28.6\%) groups.

Qualitative observations suggest that participants who referred to the documentation more often issued parameter-specific commands (e.g., directly referencing field names or value ranges), whereas others tended to rely on exploratory or descriptive instructions (e.g., ``make it harder'' or ``increase enemies''). This pattern indicates that participants engaged with the system either as a parameterized configuration space or as an open-ended interactive system, depending on how they accessed and used explicit system knowledge.

\section{Discussion}
In the present play-driven editing system, players modified gameplay through natural-language interaction with generally positive experiences and moderate workload, and no reliable differences were observed across programming-experience groups. This suggests that, within this context, natural-language editing functioned as an accessible interaction format that did not strongly depend on prior programming expertise.

Analysis of editing logs further indicated that different types of edits were associated with distinct experiential tendencies. PCA revealed interpretable editing dimensions, and trial-level analyses suggested that edits focusing on immediately perceptible parameters tended to support usability, whereas edits affecting deeper gameplay structures were more closely associated with enjoyment. Qualitative findings further contextualized these patterns: Exploratory Editing emphasized unpredictability and perceptual novelty, aligning with perceptible parameter changes; Structural Editing reflected goal-driven modification associated with deeper gameplay transformation; and Iterative Tuning highlighted a control-oriented interaction style focused on refinement and parameter understanding. Together, these results suggest that editing behaviour during play-driven interaction can follow distinct strategies that shape experiential outcomes.

Taken together, the findings indicate that play-driven game editing through voice-based natural-language interaction can function as a viable creation approach within the present system. Participants were able to modify gameplay during play regardless of programming experience, and the moderate workload scores suggest that such editing can be integrated into active gameplay without excessive cognitive burden. Rather than demonstrating a general advantage of natural-language interfaces, these results highlight the feasibility of voice-driven editing in a parameter-based arcade game context.

The documentation reference patterns reported in the Results further support the reset-based design choice described in the System section. Participants with programming experience consulted the configuration reference more frequently (83.3\%) than those without (Amateur: 50.0\%, None-EX: 28.6\%), and those who did so tended to issue parameter-specific commands rather than exploratory ones. This variation suggests that participants without technical backgrounds were not engaging with the system as a configuration interface, but rather issuing natural-language requests and reading results through gameplay behavior. This pattern is consistent with research on the abstraction gap between end-user intent and formal system representations~\cite{jiang2022whatwants}.

The qualitative themes identified in this study suggest that participants engaged with the system in ways that extend beyond task-oriented parameter adjustment. In Structural Editing, participants pursued specific aesthetic or gameplay goals --- such as approximating a bullet-hell experience (P6) or creating a psychedelic audiovisual style (P7) --- and several described the process as ``creating my own new game'' (P16). In Exploratory Editing, participants reported finding value in the unpredictability of AI responses, framing editing as a process of discovering what the system could do. These two orientations — directing the AI toward a specific goal versus treating its outputs as an open-ended creative resource — reflect distinctions described in prior work on human--AI co-creation~\cite{guzdial2019moraimaker}, though the present study did not measure these dimensions directly. Future work examining the creative dimensions of play-driven editing more explicitly would help clarify the extent to which such systems support expressive, rather than purely functional, interaction.

However, several limitations should be considered. The system relied on a structured configuration format and a retro arcade game with relatively simple rules, which may have supported stable play-driven modification. It remains unclear whether similar accessibility and workload characteristics would hold for more complex games, alternative genres, or editing scenarios involving richer rule structures or larger configuration spaces. Future work should investigate how voice-based editing scales to broader game design contexts and more complex creative tasks.

The current study is limited by its sample size of 21 participants (approximately 6--8 per programming experience group). While this sample is sufficient for exploratory analyses, the statistical power of pairwise Wilcoxon tests at these group sizes is limited, particularly for detecting small effects. The absence of statistically reliable group differences should therefore be interpreted with caution rather than as evidence of equivalence. Observed effect sizes were consistently small ($|r|$ ranged from 0.02 to 0.17, with 95\% CIs spanning zero), suggesting no practically meaningful differences across groups.

The study did not include a comparison condition, so the specific benefits of play-driven voice-based editing relative to alternative workflows cannot be directly assessed. Comparative evaluation remains a direction for future work.

It remains unclear whether the findings generalize to games with more complex rule structures, richer configuration spaces, or less universally familiar mechanics. Future work should investigate how play-driven voice editing performs across a broader range of game genres and design contexts.

\section{Conclusion}
This work presented an in-play game editing system that allows players to modify a retro arcade-style game through voice-based natural-language interaction during play. Participants were able to change gameplay parameters during play with generally positive experiences and moderate workload, and editing did not strongly depend on prior programming experience. Analysis of editing logs showed that different types of edits were associated with different experiential tendencies: adjustments to immediately perceptible parameters were linked to higher usability, whereas edits affecting core gameplay structures were more closely associated with enjoyment. Post-session reflections further revealed distinct editing strategies, including exploratory experimentation, goal-driven structural modification, and iterative parameter tuning. Together, these findings demonstrate that voice-driven editing can support accessible, in-play human -- AI co-creation within a structured configuration-based game environment.

\section*{Acknowledgements}
The hardware used in this study was funded through the University of Tsukuba Digital Nature Group 10th Anniversary Exhibition Project.



\bibliography{bibtex_main}
\bibliographystyle{unsrt}



%
%
%
%

%
%
\newpage
\vspace{5truemm}

\noindent
\textbf{Maya Grace Torii}
\vspace{3truemm} \\ 
\includegraphics[width=25truemm]{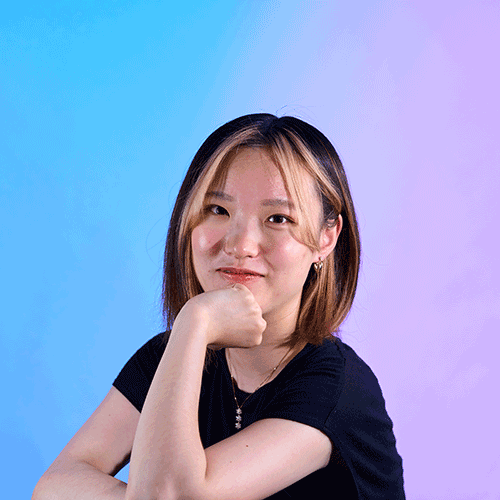} \\
Maya Grace Torii is a doctoral student at the Graduate School of Comprehensive Human Sciences, University of Tsukuba, where she joined in 2022. She received her M.S.\ degree from the University of Tsukuba in 2022 and her B.A.\ from International Christian University in 2020. Her research interests include human--AI collaborative design, play-driven interaction, and care-oriented human--computer interaction. She is a member of ACM, IEEE, IPSJ, and the Society for Nursing Engineering and Science.

\vspace{5truemm}

\noindent
\textbf{Takahito Murakami}
\vspace{3truemm} \\
\includegraphics[width=25truemm]{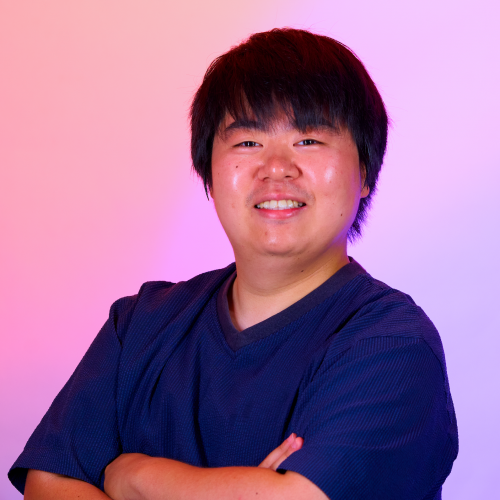} \\
Takahito Murakami is a Ph.D.\ candidate in Informatics at the University of Tsukuba and a JSPS Research Fellow (DC1). His research lies at the intersection of human--computer interaction, computer graphics, and digital fabrication, focusing on interactive and generative systems inspired by biological forms and behavioral principles. His current research explores how animal behaviors and collective biological dynamics can inform computational systems that support human creativity and design exploration. He is a member of ACM, IEEE (Robotics and Automation Society), ISAL, IPSJ, the Society for Nursing Engineering and Science, and the Japan Ethological Society.

\vspace{5truemm}
\newpage
\noindent
\textbf{Yoichi Ochiai}
\vspace{3truemm} \\
\includegraphics[width=25truemm]{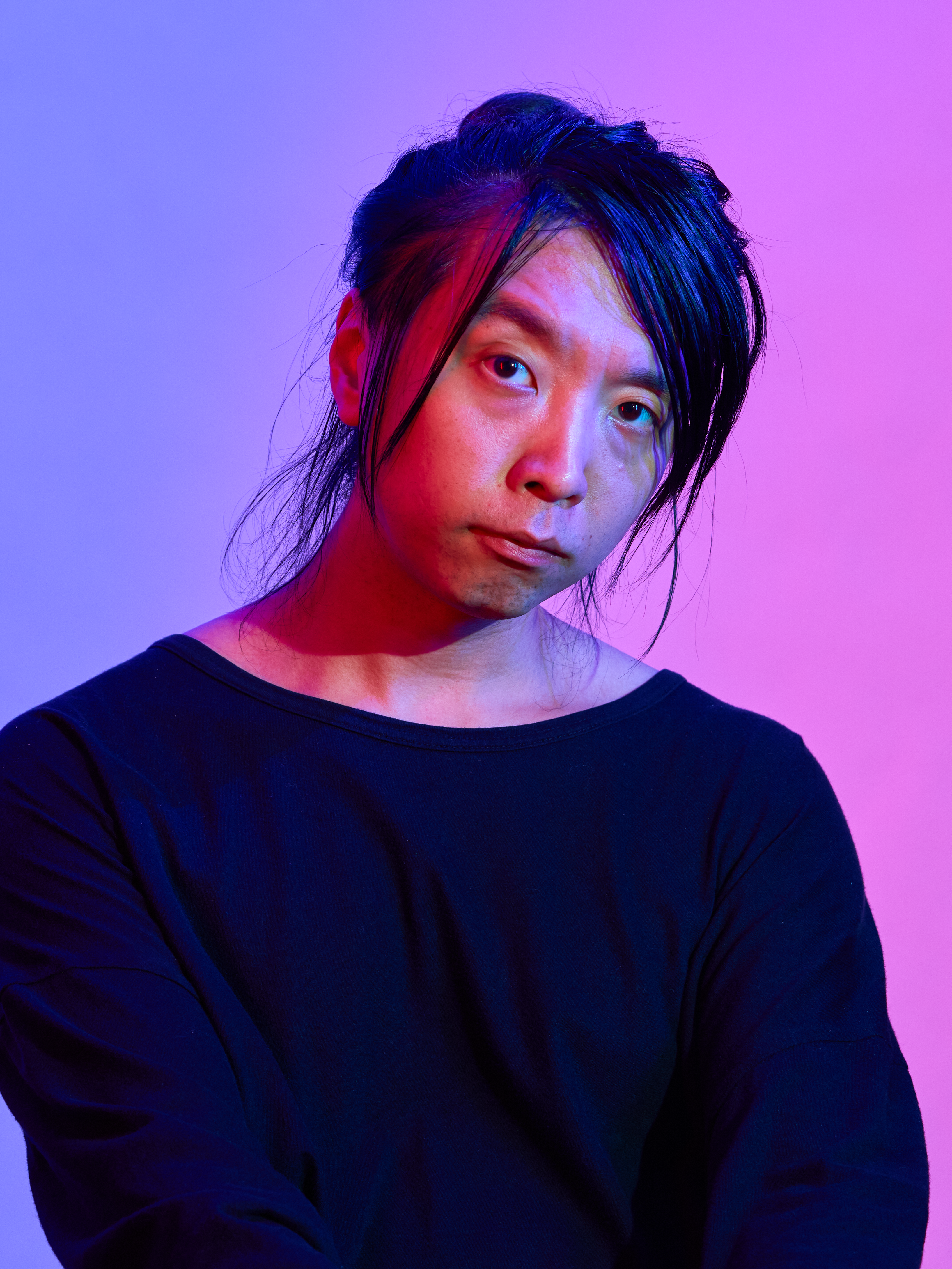} \\
Yoichi Ochiai completed his doctoral program early at the Graduate School
of Interdisciplinary Information Studies, The University of Tokyo in 2015,
receiving his Ph.D.\ in Interdisciplinary Information Studies. He was
appointed as Assistant Professor at the University of Tsukuba in the same
year, and founded Pixie Dust Technologies, Inc., serving as CEO. In 2017,
he established the Strategic Research Platform towards Digital Nature at
the University of Tsukuba. Since 2020, he has been Director of the Digital
Nature Development Research Center, University of Tsukuba, where he is
currently a Professor. He has received numerous awards including the
Minister of Education Award for Science and Technology (2023, 2025),
World Economic Forum Young Global Leader (2022), and Apollo Magazine
40 Under 40 Art and Tech (2021). He is a member of ACM, IPSJ, and VRSJ.

\end{document}